\documentclass[11pt]{article}
\usepackage{amsmath, amssymb}
\usepackage{graphicx}
\usepackage{hyperref}
\usepackage{cite}
\usepackage{xcolor}
\usepackage{fancyvrb}

\input{code/pygments_style.tex}

\usepackage{chngcntr}
\counterwithout{figure}{section}

\newcommand{\icode}[1]{\texttt{\color{blue}\small #1}}
\newcommand{\capt}[2]{\vspace{-27pt}\caption{\emph{#2}}\label{#1}}
\newcommand{\clref}[1]{\emph{\autoref{#1}}}

\title{ZjsComponent: A Pragmatic Approach to Modular, Reusable UI Fragments for Web Development}
\author{Lelanthran Manickum \\
        Rundata Systems, Johannesburg, South Africa \\
        \texttt{lelanthran@gmail.com}}

\date{\today}

\begin{document}

\maketitle

\begin{abstract}
In this paper, I present \emph{ZjsComponent}, a lightweight and
framework-agnostic web component designed for creating modular, reusable UI
elements with minimal developer overhead. \emph{ZjsComponent} is an example
implementation of an approach to creating components and object instances
that can be used purely from HTML.

Unlike traditional approaches to components, the approach implemented by
\emph{ZjsComponent} does not require build-steps, transpiling,
pre-compilation, any specific ecosystem or any other dependency. All that
is required is that the browser can load and execute Javascript as needed
by Web Components.

\emph{ZjsComponent} allows dynamic loading and isolation of HTML+JS fragments,
offering developers a simple way to build reusable interfaces with ease.
This approach is dependency-free, provides significant DOM and code isolation,
and supports simple lifecycle hooks as well as traditional methods expected of
an instance of a class.  I evaluate its functionality through practical
examples and discuss its potential impact on web development.
\end{abstract}

\section{Introduction}
Web development has evolved significantly over the last decade, with a variety
of frameworks providing powerful tools for building complex applications.
However, many of these frameworks introduce complexity and
overhead, requiring developers to manage intricate configurations and adopt
unfamiliar paradigms. In contrast, \emph{ZjsComponent}
is designed to provide a simple, pragmatic approach for creating modular and
reusable UI fragments with minimal setup. By leveraging dynamic HTML+JS
fragment loading, \emph{ZjsComponent} allows developers to compose components
without relying on any framework. This paper discusses the goal, design,
implementation and potential use cases of \emph{ZjsComponent}.

\emph{ZjsComponent} can be downloaded from Zenodo\cite{ZjsComponent-v1.0.0-rc1}.

\section{Related Work}
Many component-based frameworks
exist\cite{react}\cite{angular}\cite{vue}\cite{solidjs}\cite{alpinejs}\cite{preact}
which allow developers to build reusable UI components. These frameworks
provide rich features, such as two-way data binding, state management, and
reactivity.  However, these features come at the cost of increased complexity
and steep learning curves.

\emph{Web Components}\cite{webcomponents}\cite{Shah_2023} offer a web-native,
standards-based approach to building reusable UI elements but often lack
flexibility in loading external scripts and handling dynamic content.
Approaches exist \cite{Fouquet_2022} to remove \emph{all} Javascript from the
document, but are limited to only the interactivity provided by CSS alone.

\emph{ZjsComponent} distinguishes itself by focusing on simplicity and
minimalism by:
\begin{itemize}
    \item {Offering a low-overhead alternative that still supports powerful
        use cases,}
    \item {Minimal Javascript overhead (fewer than 100 lines in the
        suggested implementation\cite{ZjsComponent-v1.0.0-rc1}),}
    \item {Encourages an HTML-first approach to interactive components.}
\end{itemize}

While \emph{ZjsComponent} is not necessarily a disappearing
framework\cite{Veps_l_inen_2023}, the end result of HTML and plain Javascript
is almost indistinuishable from a disappearing framework.

\section{Goal: Reducing Cognitive Burden to Zero}
Modern web development often requires developers to learn complex tools like
\emph{npm}, \emph{webpack}, \emph{TypeScript}, \emph{JSX}, and that's before
even learning the patterns and API specific to the framework in question.
In practice, there's many more tools to learn before one can be
productive in the environment.

This creates significant cognitive overhead, especially for newcomers and
teams who just want to build functional applications.

\emph{ZjsComponent} aims to eliminate this complexity by reducing the
cognitive burden to near zero.
\begin{itemize}
   \item[\textbf{Creation}] Requires only HTML tags and Javascript functions.
   \item[\textbf{Import}] Requires only \icode{<script src=...>}.
   \item[\textbf{Usage}] Requires only \icode{<zjs-component remote-src=...>}.
\end{itemize}
Other than a text editor, nothing more is
needed to author documents with reusable components. Each component is an
isolated fragment of valid HTML.

The practical result of this approach supports delivering and using Javascript
objects directly from HTML. The benefits of inserting components into a
document using nothing more than \texttt{<zjs-component>} tags are numerous:

\begin{itemize}
   \item \textbf{Focus on Functionality, Not Configuration}:
      \emph{ZjsComponent} eliminates the need for build tools and
      dependencies, allowing developers to focus on building features rather
      than dealing with configuration and setup.
   \item \textbf{Lower Barriers to Entry}: By removing the need to learn
      complex frameworks or tools, \emph{ZjsComponent} makes it easier for
      developers, especially beginners, to create reusable components with
      minimal effort.
   \item \textbf{Improved Productivity}: With no complicated build steps or
      dependencies, developers can quickly start building and integrating
      components, leading to faster development cycles.
   \item \textbf{Simplified Maintenance}: \emph{ZjsComponent}'s lack of
      external dependencies reduces the risk of version conflicts and
      compatibility issues, simplifying long-term maintenance.
   \item \textbf{Reusability and Flexibility}: Components created with
      \emph{ZjsComponent} can be easily reused across projects with just a
      simple \icode{<script src="...">} tag, streamlining development and
      reducing duplication of effort.
\end{itemize}

By focusing on simplicity and eliminating unnecessary complexity,
\emph{ZjsComponent} provides a practical approach to building reusable
components, making it accessible to developers of all experience levels.

\section{Design and Implementation}
\subsection{Core Design}
\emph{ZjsComponent} is a custom HTML element\cite{Shah_2023} that allows
clients to load HTML+JS fragments dynamically using a \icode{remote-src=}
attribute. When a \icode{<zjs-component>} is added to the DOM, it fetches the
content from the specified \icode{remote-src=} attribute and inserts the HTML
fragment into the component's DOM. The scripts within the fragment are
executed in isolation, ensuring that there are no side effects on the parent
document.  \emph{ZjsComponent} supports two lifecycle hooks:
\begin{itemize}
   \item \icode{onConnected()}: Called when a fragment is inserted into the
      DOM.
   \item \icode{onDisconnected()}: Called when a fragment is removed from the
      DOM.
\end{itemize}

The result is a \emph{ZjsComponent} that is, in practice, just another
Javascript object with methods that can be be called, albeit one that can
be used from within HTML only.

The full source code for the implementation can be found at
\href{https://TODO.todo.com/todo}{TODO}.
\subsection{Dynamic Loading and Isolation}
The key feature of \emph{ZjsComponent} is its ability to load HTML+JS
fragments dynamically. This is achieved using the \icode{remote-src=}
attribute, which points to an external file containing both HTML and script
content. Upon loading, the fragment is inserted into the component, with any
scripts executed in a closure to avoid polluting the global namespace.

\subsection{Method Invocation}
From within Javascript, a method on any \icode{ZjsComponent} can be invoked in
the normal way using dot syntax, for example \icode{myComponent.foo()}.

To interact with a loaded component, \emph{ZjsComponent} also provides a
static \icode{send()} method. This allows developers to invoke methods defined
within the component's script, such as updating the UI or triggering actions.
The \icode{send()} method simplifies communication with the component, making
it easy to integrate into larger applications without complex event-handling
systems.

\section{Examples and Use Cases}
\subsection{Client-Side Includes with \emph{ZjsComponent}}
In its most basic form, \emph{ZjsComponent} can be used merely for
\emph{client-side includes} i.e. including a fragment of HTML within the page,
as is normally required for headers, footers and other repeated static
content.

\clref{fig:client-side-includes-1} shows an example of how to use
\emph{ZjsComponent} to inject a header onto a page.  First the component is
written as a fragment of HTML.

\begin{figure}[htbp]
   \centering
   
  \begingroup
     \RecustomVerbatimEnvironment{Verbatim}{Verbatim}{
       frame=lines,
       framesep=5pt,
       fontsize=\footnotesize,
       numbers=left,
       commandchars=\\\{\}
     }%
    \input{code/a-sample-html-fragment.tex}
  \endgroup

   \capt{fig:client-side-includes-1}{A sample HTML fragment.}
\end{figure}

At the call-site, the \emph{ZjsComponent} is used as shown in
\clref{fig:client-side-includes-2}.

\begin{figure}[htbp]
   \centering
   
  \begingroup
     \RecustomVerbatimEnvironment{Verbatim}{Verbatim}{
       frame=lines,
       framesep=5pt,
       fontsize=\footnotesize,
       numbers=left,
       commandchars=\\\{\}
     }%
    \input{code/using-a-fragment.tex}
  \endgroup

   \capt{fig:client-side-includes-2}{Using a fragment.}
\end{figure}

In \clref{fig:client-side-includes-2}, the content of \icode{content.html}
will be fetched and rendered inside the \icode{<zjs-component>} element. The
\icode{display} style attribute can be set using the attribute
\icode{display=<value>}. This allows the user of the component to decide if
it must be displayed inline or as a block, etc.

\subsection{Passing parameters}
\clref{fig:parameters-1} is an example of using a \icode{<zjs-component>} in
HTML that allows the user of the specified component to pass in parameters
using attributes.

\begin{figure}[htbp]
  \centering
   
  \begingroup
     \RecustomVerbatimEnvironment{Verbatim}{Verbatim}{
       frame=lines,
       framesep=5pt,
       fontsize=\footnotesize,
       numbers=left,
       commandchars=\\\{\}
     }%
    \input{code/passing-parameters-to-a-fragment.tex}
  \endgroup

  \capt{fig:parameters-1}{Passing parameters to a fragment.}
\end{figure}

This will load the component from the \icode{component/hello.zjsc} file,
making the \icode{greeting=} and \icode{name=} attributes available to all the
methods within the component.

The component itself must be written to use those attributes, if they exist,
as shown in \clref{fig:parameters-2}.

\begin{figure}[htbp]
  \centering
   
  \begingroup
     \RecustomVerbatimEnvironment{Verbatim}{Verbatim}{
       frame=lines,
       framesep=5pt,
       fontsize=\footnotesize,
       numbers=left,
       commandchars=\\\{\}
     }%
    \input{code/a-fragment-accepting-parameters.tex}
  \endgroup

   \capt{fig:parameters-2}{A fragment accepting parameters.}
\end{figure}

Special attention is drawn to the use of the \icode{this} keyword in the
\clref{fig:parameters-2} code snippet: the component is a proper object and
can be used as such.

\subsection{Parameterised Example}
Methods are invoked on a \emph{ZjsComponent} by \textbf{sending} the instance
a message
using the static \icode{ZjsComponent.send()} function. This allows
direct usage from HTML elements within an HTML page without needing a separate
\icode{<script>} element containing boilerplate to locate and invoke the
correct instance of a \emph{ZjsComponent}.

\clref{fig:substantial-1} is a more substantial example containing a
\icode{<button>} that is used to update the component's display.

\begin{figure}[htbp]
  \centering
   
  \begingroup
     \RecustomVerbatimEnvironment{Verbatim}{Verbatim}{
       frame=lines,
       framesep=5pt,
       fontsize=\footnotesize,
       numbers=left,
       commandchars=\\\{\}
     }%
    \input{code/a-more-complex-usage-calling-methods-on-zjscomponents.tex}
  \endgroup

  \capt{fig:substantial-1}{A more complex usage calling methods on \emph{ZjsComponents}.}
\end{figure}

The updated \icode{hello.zjsc} component is shown in
\clref{fig:substantial-2}.  In the \clref{fig:substantial-2} snippet the
first parameter to \icode{ZjsComponent.send()} is a query selector string.
However, the first component may also be a value of type \icode{ZjsComponent}.
While it is possible to call \icode{ZjsComponent.send()} with a reference to a
\icode{ZjsComponent} as the first argument, it's more efficient and
understandable to simply do \icode{myVar.updateName(...)} when \icode{myVar}
is a variable holding a value of type \icode{ZjsComponent}.

\begin{figure}[htbp]
  \centering
   
  \begingroup
     \RecustomVerbatimEnvironment{Verbatim}{Verbatim}{
       frame=lines,
       framesep=5pt,
       fontsize=\footnotesize,
       numbers=left,
       commandchars=\\\{\}
     }%
    \input{code/the-zjscomponent-on-which-methods-are-called.tex}
  \endgroup

   \capt{fig:substantial-2}{The \emph{ZjsComponent} on which methods are called.}
\end{figure}

\subsection{Self-referential Example}
The static method \icode{ZjsComponent.send()} is a little more flexible,
supporting brevity when used inline within HTML. The first parameter can be
any one of the following:
\begin{itemize}
   \item A query selector string, which is used as the first parameter in a
      call to
      \icode{querySelector()} to locate the \emph{ZjsComponent} in
      the document.
   \item An object of type \icode{ZjsComponent}.
   \item Any element which is a descendent of a \icode{ZjsComponent}; in this
      case the closest \icode{ZjsComponent} element is located and the
      \icode{ZjsComponent.send()} method of that object is invoked.
\end{itemize}

Dispatching an invocation to the closest \emph{ZjsComponent} ancestor
allows fragments to easily reference the containing \emph{ZjsComponent}
object, even from within inline HTML.

The inline HTML fragment in \clref{fig:self-1} is easily able to invoke
methods on the object using only HTML attributes.

\begin{figure}[htbp]
  \centering
   
  \begingroup
     \RecustomVerbatimEnvironment{Verbatim}{Verbatim}{
       frame=lines,
       framesep=5pt,
       fontsize=\footnotesize,
       numbers=left,
       commandchars=\\\{\}
     }%
    \input{code/calling-send-on-the-nearest-ancestor-of-an-element.tex}
  \endgroup

  \capt{fig:self-1}{Calling `send()' on the nearest ancestor of an element.}
\end{figure}

\section{Evaluation and Discussion}
\emph{ZjsComponent} provides a lightweight solution for building reusable UI
components using only fragments of valid HTML, \textbf{without} the need for a
full-fledged framework. Its performance is optimized for minimal overhead, and
its design encourages developers to focus on the functionality of individual
components without being encumbered by complex configuration. One of the main
benefits of \emph{ZjsComponent} is its simplicity; it can be easily integrated
into existing projects without requiring any changes to the existing
architecture.

\subsection{Anticipated arguments}
I anticipate persuasive arguments against \emph{ZjsComponent}.
\subsubsection{Functionality already available}
\textbf{Existing frameworks already provide this functionality, and more.}
While existing frameworks do indeed provide the facilities provided by
\emph{ZjsComponent} the use-case and intended audience of \emph{ZjsComponent}
has little overlap with the use-cases and intended audience of many of the
existing frameworks. Usage of \emph{ZjsComponent} requires an author to
know only the \icode{<zjs-component>} tag. Creating a new \emph{ZjsComponent}
requires only knowledge of HTML and Javascript. There are no build-steps,
technology stacks or reams of documents to read before one can produce or use
reusable HTML components.

\subsubsection{No Reactivity}
\textbf{\emph{ZjsComponent} does not provide reactivity\cite{reactivity_vue},
state management\cite{state_management_redux} or two-way
data-binding\cite{two_way_binding_angular}.}
This is true; reactivity, state-management and two-way data-binding are out of
scope for a \emph{component creation} mechanism. \emph{ZjsComponent} is not a
\emph{React} or \emph{SolidJS} replacement, it is an approach to creating
reusable components.

However, this does not rule out the
creation of \emph{ZjsComponents} that make some headway into providing this
functionality.

\subsubsection{CSS is not isolated}
\textbf{A component's \icode{style} element affects the whole
document.}
This is an intentional trade-off. Isolating the document from the
component's styles also isolates the component from the document's styles.
It's a trade-off between easily allowing components the ability to bring their
own styles and allowing components to use a site-wide style.\cite{Shah_2023}

\subsubsection{Duplicated fetching}
\textbf{Each \emph{ZjsComponent} will fetch the document specified in the
\emph{remote-src} attribute, even if many components have the same
\icode{remote-src=} attribute.}
This is superficially correct. In practice the browser would simply serve up a
cached copy of the fragment specified in the \icode{remote-src=} attribute.
Where this would be an issue is in the case that the first fetch of the
fragment is still in-flight, and thus there is no cached copy to use. I
propose a simple mitigation using a static field within \icode{ZjsComponent}
to hold promises for each \icode{fetch} attempt. The fetching would then
consist of \icode{await}ing on these promises if they exist, and populating
that field with an additional promise if it doesn't.

\section{Future Directions}
A number of currently complex framework-based solutions can be replaced by
simpler solutions using \emph{ZjsComponent}. For example, some preliminary
experimentation with front-end route management have shown a small measure
of success. It is not difficult to envision more complex components being
written as a composition of multiple simpler fragments.

\section{Conclusion}
In this paper, I present \emph{ZjsComponent}, a lightweight and simple web
component designed to provide developers with an easy-to-use, modular way to
build reusable UI elements. \emph{ZjsComponent} emphasizes practicality and
minimal overhead, allowing for rapid integration and simple component
management.  While it has known limitations the approach offers a useful
alternative to more complex frameworks.

In practice, the ability to use nothing more than plain HTML sans
build-step to produce a full and complete Javascript object can be a driver
for more powerful framework architecture for common usage.

\section{References}

\bibliographystyle{plain}
\bibliography{zjs-component.bib}

\end{document}